\documentclass[aps,prl,reprint,superscriptaddress,floatfix,longbibliography]{revtex4-2}

\usepackage{blindtext,amsmath}
\usepackage{graphicx}
\usepackage{float}
\usepackage[export]{adjustbox}

\newcommand{\gib}{g_\mathrm{IB}}
\newcommand{\mred}{m_\mathrm{r}}

\newcommand{\mb}{m}
\newcommand{\mi}{M}

\usepackage{color}

\newcommand{\dn}{\textrm{dn}}
\newcommand{\cn}{\textrm{cn}}
\newcommand{\cd}{\textrm{cd}}
\newcommand{\sn}{\textrm{sn}}
\newcommand{\am}{\textrm{am}}

\newcommand{\xit}{\bar{\xi}}

\usepackage{xcolor}
\definecolor{Cerulean}{rgb}{0.,0.59,0.835}
\definecolor{RubineRed}{rgb}{0.61,0.07,0.12}
\usepackage[colorlinks=true,citecolor=Cerulean,linkcolor=RubineRed,urlcolor=Cerulean,pdftex]{hyperref}

\begin{document}
\title{Polaron Interactions and Bipolarons in One-Dimensional Bose Gases in the Strong Coupling Regime}
\author{M.~Will}
\affiliation{Department of Physics and Research Center OPTIMAS, University of Kaiserslautern, 67663 Kaiserslautern, Germany}
\author{G. E. Astrakharchik}
\affiliation{Departament de Física, Universitat Politècnica de Catalunya, Campus Nord B4-B5, E-08034, Barcelona, Spain}
\author{M.~Fleischhauer}
\affiliation{Department of Physics and Research Center OPTIMAS, University of Kaiserslautern, 67663 Kaiserslautern, Germany}

\begin{abstract}
Bose polarons, quasi-particles composed of mobile impurities surrounded by cold Bose gas, can experience strong interactions mediated by the many-body environment and form bipolaron bound states. Here we present a detailed study of heavy polarons in a one-dimensional Bose gas by formulating a non-perturbative theory and complementing it with exact numerical simulations. We develop an analytic approach for weak boson-boson interactions and arbitrarily strong impurity-boson couplings. Our approach is based on a mean-field theory that accounts for deformations of the superfluid by the impurities and in this way minimizes quantum fluctuations. The mean-field equations are solved exactly in Born-Oppenheimer (BO) approximation leading to an analytic expression for the interaction potential of heavy polarons which is found to be in excellent agreement with quantum Monte-Carlo (QMC) results. In the strong-coupling limit the potential substantially deviates from the exponential form valid for weak coupling and has a linear shape at short distances. Taking into account the leading-order Born-Huang corrections we 
calculate bipolaron binding energies for impurity-boson mass ratios as low as 3 and find excellent agreement with QMC results.
\end{abstract}

\date{\today}

\maketitle


\paragraph{Introduction:--}
Interactions between quantum particles mediated by a many-body environment play an important role in condensed-matter physics. Examples range from the Ruderman-Kittel-Kasuya-Yodsia (RKKY) interaction of spins in a Fermi liquid \cite{Ruderman1954,Kasuya1956,Yosida1957} to Cooper pairing of electrons in a solid induced by lattice vibrations \cite{Cooper1956}. The mechanism that causes such interactions can 
also substantially modify the properties of individual impurities by forming quasi-particles. A paradigmatic example is the polaron \cite{Landau1933,Pekar1946} resulting from the electron-phonon coupling also responsible for Cooper pairing. In the strong coupling limit, impurity interaction and quasi-particle formation are strongly 
intertwined. Bipolarons are suspected to be essential for high-temperature superconductivity \cite{Alexandrov1992,Mott1993,Alexandrov1994}. They are important for 
the electric conductivity of polymers \cite{Bredas1985,Glenis1993,Bussac1993,Fernandes2005,Zozoulenko2019} or organic magneto-resistance \cite{Bobbert2007}. Their understanding is 
one of the key questions of many-body physics.

In recent years neutral atoms immersed in degenerate quantum gases have become versatile experimental platforms for accessing polaron physics in novel regimes and with an unprecedented degree of control 
\cite{swa09,Zhang2012,kohstall_metastability_2012,koschorreck_attractive_2012,Scazza2016,cetina_decoherence_2015,cetina_2016,
Schirotzek2009,Kohstall2012,Koschorreck2012,Catani2012,Hohmann2015,Hu,Jorgensen2016,Hu,Scazza2017,Yan2020}. Length and energy scales are 
very different from solids and can be resolved and manipulated much easier. Most importantly polarons can be studied out of equilibrium with the prospect of engineering their properties beyond what is possible in equilibrium. One-dimensional (1D) gases are of particular relevance as they show pronounced quantum effects and powerful tools are available for their theoretical description. It is possible to tune the impurity-bath interaction all the way through weak to strong coupling, e.g. by employing Feshbach and confinement induced resonances (CIR) \cite{Olshanii98}. Contrarily to higher dimensions, the system remains stable even for infinite coupling since three-body losses are greatly suppressed.
\begin{figure}[htb]
\centering
\includegraphics[width=0.35\textwidth]{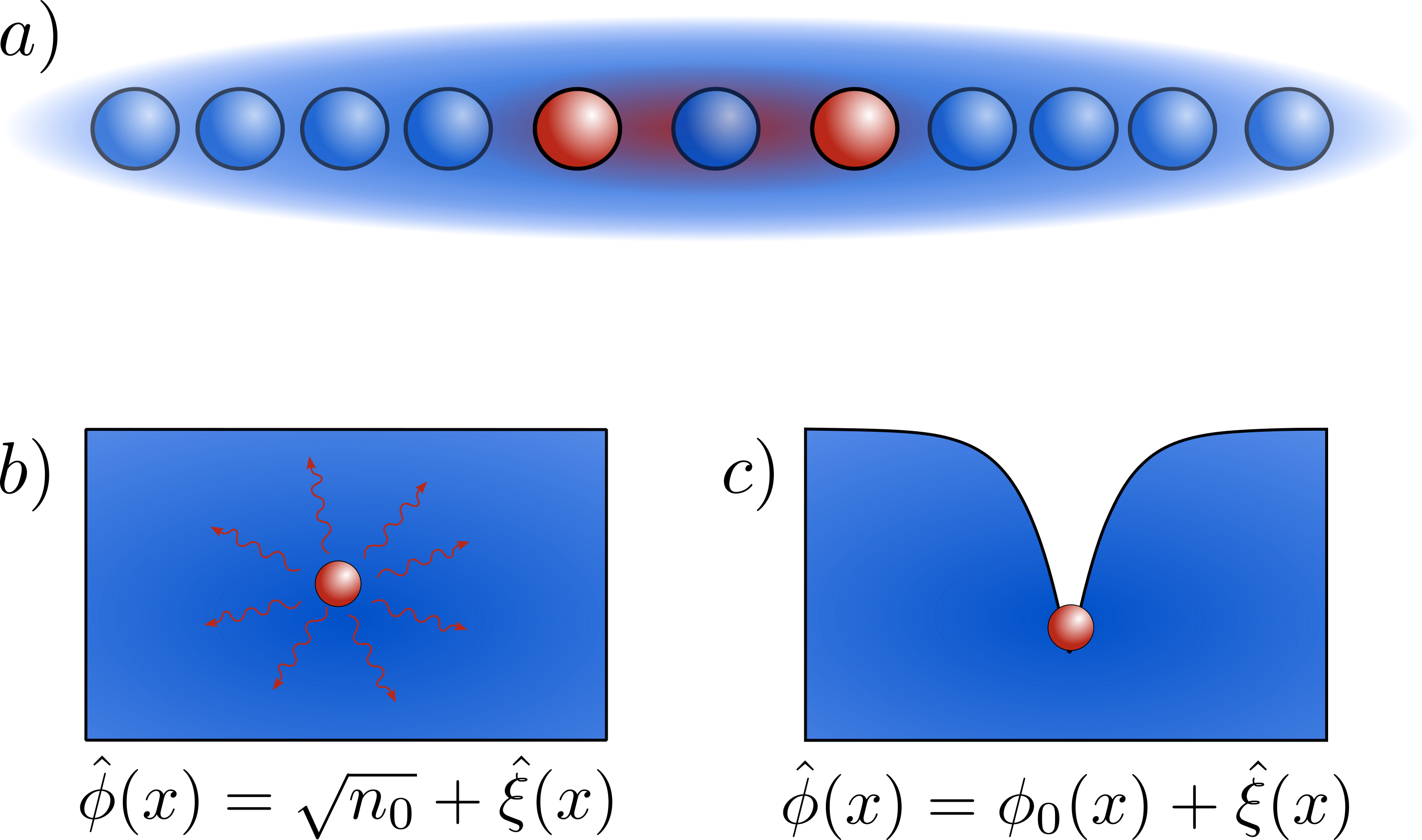}
\caption{(a) Sketch of a bipolaron composed of two impurities in a 1D Bose gas. (b) In the commonly used extended Fr\"ohlich model a large number of phonons are created around the impurity and phonon-phonon interactions need to be taken into account. (c) In a description based on a deformed condensate this can be avoided.}
\label{fig:illustration}
\end{figure}
Polaron interactions have so far mostly been studied in regimes where the mediated interaction 
between them 
is weak.
A perturbative treatment yields an exponential (1D) or Yukawa (3D) potential between two impurities with the characteristic length scale set by the healing length $\xi$ \cite{pethick2002book,Klein2005,Recati2005,Naidon2018}. 
A universal low-energy theory of mobile impurities in one dimension has been developed in Ref.~\cite{Schecter2014}, restricted to particle separations much larger than $\xi$ where the interaction is weak. A unified treatment for all distances, but for immobile impurities and small impurity-boson couplings has been given in Ref.~\cite{Reichert2019PRB,Reichert2019NJP}. While QMC methods have been used to obtain polaron properties in a non-perturbative manner \cite{Astrakharchik2013,Grusdt2017b,Tylutki2017,Mistakidis2019,Ardila2020,Astrakharchik2020ionic} and there are numerical mean-field
studies in trapped systems extending into the non-perturbative regime \cite{Dehkharghani2018}, analytic approaches have been restricted to weak polaron-polaron couplings or non-interacting host gases \cite{Flicker1967,Huber2019,Dean2021}. The first attempt to strong polaron couplings in interacting gases has been made only recently by using a scattering-matrix expansion \cite{CamachoGuardian2018}. The authors predict a deviation from the 3D Yukawa potential in 
agreement with QMC simulations but with 
some notable quantitative differences.

Here we develop an analytic theory of polaron interactions in 1D Bose gases for arbitrary strength of the impurity-boson coupling, see Figure~\ref{fig:illustration}a for a sketch. A common 
description of the Bose-polaron takes into account a coupling of the impurity only to Bogoliubov phonons \cite{Shchadilova2016,Naidon2018}, see Fig.~\ref{fig:illustration}b. This extended Fr\"ohlich model is however not adequate for strong coupling, $\gib\gg g$, even if the boson-boson interaction itself is weak, since the impurity generates a high-density cloud of phonons around it and phonon-phonon interactions can no longer be neglected. Here we use a different approach that accounts for deformation of the superfluid by the impurities, see Fig.~\ref{fig:illustration}c
\cite{Volosniev2017,Mistakidis2019,Jager2019}.
As shown in Ref.~\cite{Jager2019} and elucidated in the Supplemental Material this approach 
minimizes quantum fluctuations and
leads to highly accurate predictions for single-polaron properties already on the mean-field level, precise enough to differentiate finite-size effects. Employing this approach, we develop a mean-field theory of bipolarons assuming a weakly interacting condensate and moderately heavy impurities and verify the semi-analytic predictions with QMC results. 

\paragraph{Model:--}

We consider two impurities of equal mass $M$ in a 1D gas of bosons of mass $m<M$. We assume contact impurity-boson interactions with coupling strength $\gib$ which can be repulsive, $\gib>0$, or attractive, $\gib<0$. We disregard a direct interaction between the impurities.
Introducing center-of-mass (COM) and relative impurity coordinates $\hat R, \hat r$ and momenta $\hat P, \hat p$, the Hamiltonian reads ($\hbar =1$)
\begin{eqnarray}
\hat{H} &=& \frac{\hat{P}^2+ 4\hat{p}^2}{4 \mi} + \!\int\!\! dx \, \hat{\Phi}^\dagger (x) \, \Bigg\{ \frac{-1}{2 \mb} \partial_x^2 \, + \, \frac{g}{2} \hat{\Phi}^\dagger(x)\hat{\Phi}(x) \label{H_start} \\
	 &-&\mu + \gib \, \Big[ \delta\left(x- \hat{R}-\frac{\hat{r}}{2}\right) +\delta\left(x-\hat{R}+ \frac{\hat{r}}{2}\right) \Big]\Bigg\} \hat{\Phi}(x)
	, \notag 
	\end{eqnarray}
Here $\mu$ is the chemical potential of the gas, which in mean-field approximation is $\mu = g n_0$, with $n_0$ being the linear density far away from both impurities. In the thermodynamic limit $n_0$ converges to the mean density $n=N/L$. The interaction between the bosons of strength $g$ is assumed to be weak so that a Bogoliubov approximation applies, i.e. the healing length $\xi=1/\sqrt{2 m \mu}$ \cite{Pethick2008} is large compared to the mean inter-particle distance $1/n$. This regime is characterized by a small Lieb-Liniger parameter $\gamma=m g/n$ \cite{Lieb1963}. The dependence of the COM coordinate can be eliminated using a Lee-Low-Pines (LLP) transformation \cite{Lee1953} $\hat{U} = \exp \left(-i \hat{R} \hat{P}_{\text{B}} \right)$, where $\hat{P}_{\text{B}} = -i \int dx \, \hat{\Phi}^\dagger (x) \partial_x \hat{\Phi} (x)$ is the total momentum of the Bose gas: 
\begin{align}
&\hat{H}_\textrm{LLP} = \frac{:\!\!\left(P-\hat{P}_{\text{B}}\right)^2\!\! :+4\hat{p}^2}{4\mi} 
 + \int \!\! dx \, \hat{\Phi}^\dagger (x) \, \Bigg\{\frac{-1}{2 \mred} \partial_x^2 \, \label{HLLP}\\
 &\, -\mu+\frac{g}{2} \hat{\Phi}^\dagger \hat{\Phi} + \gib \, \Big[ \delta\left(x- \frac{\hat{r}}{2}\right)
 +\delta\left(x+ \frac{\hat{r}}{2}\right) \Big] \Bigg\} \hat{\Phi}(x),\notag
\end{align}
where $:\, :$ denotes normal ordering, i.e. interchanging all creation operators to the left and annihilation to the right, and $\mred = 2Mm/(2M+m)$ is the reduced mass. $\hat{P}$, which previously was the COM momentum of the two impurities, is in the new frame the total momentum of the system. It is a constant of motion which can be replaced by a c-number $P$, and we set $P=0$. Note that the LLP transformation is needed for any $M<\infty$ even if one considers an impurity at rest.

\paragraph{Bipolaron of heavy impurities:--}
Differently from the single polaron case, the LLP transformation does not remove the impurity coordinates entirely. To this end we apply a Born-Oppenheimer (BO) approximation, valid for $\mi\gg \mb$, where the kinetic energy of the relative motion is neglected and one can replace $\hat r$ by a c-number $r$. This turns $\hat H_\textrm{LLP}$ into a pure boson Hamiltonian.

In the following we determine the ground state of~\eqref{HLLP} for a weakly interacting gas which amounts to assume small quantum fluctuations $\hat\xi(x)$ on top of the mean-field solution $\phi_0(x)$ of Eq.~\eqref{HLLP}, $\hat \phi(x) =\phi_0(x) +\hat \xi(x)$. Note that this differs from the common approach, where a small-fluctuation expansion is applied in the absence of the impurities first. In contrast we take the back-action of the impurity into account already at the mean-field level. As shown in Ref.~\cite{Jager2019} this (i) leads to modified Bogoliubov phonons, coinciding with the standard ones only in the long-wavelength limit $k\xi \gg 1$, and (ii) minimizes their generation by the impurity, see Fig.~\ref{fig:illustration}. The smallness of quantum fluctuations allows us to ignore them altogether when considering the mediated impurity-impurity interaction at distances of the order of a few re-scaled healing lengths, $\bar{\xi}= \sqrt{m/\mred}\xi$. Only at large separations quantum fluctuations become relevant. They are responsible for weak Casimir-type interactions scaling as $1/r^3$ \cite{Schecter2014,Reichert2019PRB,Reichert2019NJP} for finite $\gib$ and $1/r^2$ or $1/r$ if either one or both of the static impurities have infinitely strong coupling \cite{Reichert2019,Petkovic2020}. We will not consider these contributions here but show a posteriori that the corrections are small on absolute scale.

The mean-field solutions of~\eqref{HLLP} can be obtained analytically in the BO limit, see Supplemental Material. In particular one finds for the interaction potential between two impurities
\begin{align}
	& V(r)
	= g n_0^2 r \left(\frac{1}{2}- \frac{4+2\nu}{3(\nu+1)^2} \right)+	
	\frac{4}{3} \frac{g n_0^2 \xit}{\sqrt{1+\nu}} \Bigg\{
\sqrt{2\nu+2}
	 \notag\\
	 &+ 2 \, \text{E} \big(\am(u,\nu),\nu\big) - \frac{\sqrt{\tilde{v}}^3}{1+v} \cd(u,\nu)^{\pm 3} \left[1+\sqrt{\tilde{v}} \; \sn(u,\nu) \right]
	\notag \\
	& - \sqrt{\tilde{\nu}} \;	\cd(u,\nu)^{ \pm 1} \left[ \frac{3}{2} + \frac{1+\nu+\tilde{\nu}}{1+\nu}\sqrt{\tilde{\nu}} \; \sn(u,\nu)
	\right]
	\Bigg\}, \label{eq:V_r}
\end{align}
where 
$u ={r}/(2\bar{\xi} \sqrt{1+\nu})$ is a normalized distance, and the upper (lower) sign stay for repulsive (attractive) impurity-boson interaction, $\text{E}(x,\nu)$ is the incomplete elliptic integral of the second kind, $\cd(x,\nu)$ and $\sn(x,\nu)$ are Jacobi elliptic functions and $\am(x,\nu)$ is the amplitude of these functions \cite{Lawden1989}. The dimensionless parameter $\nu = \nu(r,\eta)$ with $|\nu|<1$ is given implicitly by 
\begin{align}
2 \frac{|\eta|}{n_0 \xit} \frac{\sqrt{\tilde{\nu}(\nu+1)}}{(1-\nu)} \; &\cn \left(u, \nu \right)\, \dn \left(u, \nu \right)= \left[ 1 + \sqrt{\tilde{\nu}} \,\sn \left(u, \nu \right) \right]^2, 
\nonumber
\end{align}
involving the Jacobi elliptic sn, cn, and dn functions and $\eta = \gib / g$. Here $\tilde{\nu} = \nu$ for $\eta>0$ and $\tilde{\nu} = 1$ for $\eta<0$. In general, this equation has several solutions, however the physically relevant one is that with the largest $\nu$.

\begin{figure}[htb]
\centering
\includegraphics[width=0.48\textwidth]{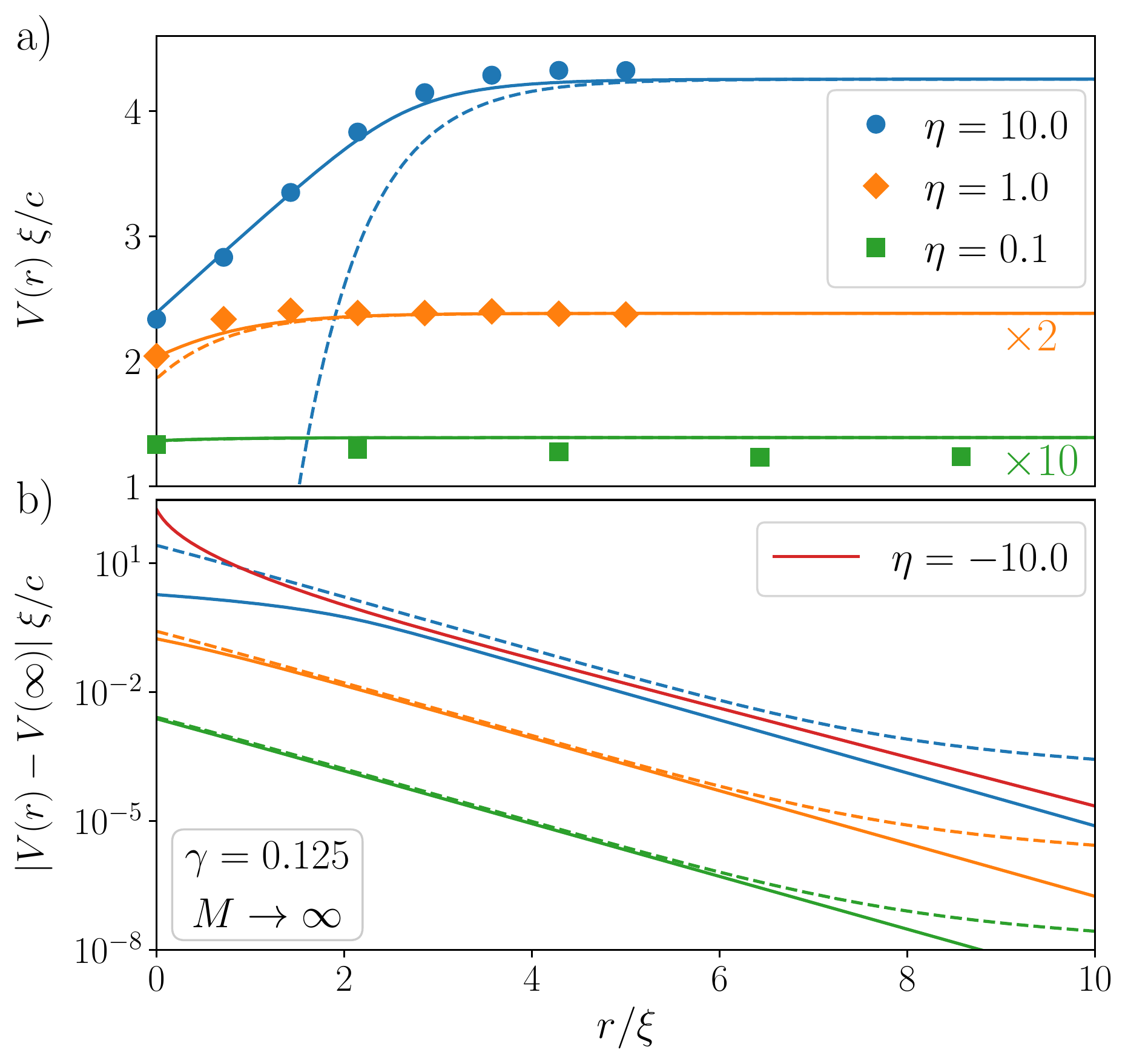}
\caption{Effective impurity-impurity interaction as function of distance in units of $\xit=\xi$ for different interactions $\eta=g_\textrm{IB}/g$ and $M\to\infty$, where $c=\sqrt{gn_0/m}$ is the speed of sound. Solid lines represent semi-analytical approximation Eq.~\eqref{eq:V_r}, circles are QMC results (errorbars smaller than circle size) and dashed lines give  perturbative predictions from \cite{Reichert2019PRB} including Casimir-type contribution. (a) Comparison of effective potential $V(r)$ for repulsive impurity-boson interaction. The perturbative results were shifted to match our predictions at infinite distance. (b) Interaction potential on a semi-log scale. Exponential decay for weak impurity-boson couplings, $\eta \lesssim 1$, is seen as straight lines. The Casimir effect (absent in the mean-field description) results in the slower $1/r^3$ decay at $r\gtrsim 6 \xi$.
}
\label{fig:BO-potentials}
\end{figure}

Figure~\ref{fig:BO-potentials} shows examples of the effective interaction potential
$V(r)$, having a finite range defined by $\bar\xi$. The strong coupling regime is reached when 
$\eta \gtrsim n_0\xit = 1/\sqrt{2\frac{m_\textrm{r}}{m}\gamma}$, \cite{Jager2019}. In this case the impurity causes a sizable deformation of the Bose gas and $V(r)$ deviates substantially from the perturbative exponential behavior at short distances predicted in Ref.~\cite{Reichert2019NJP}. The logarithmic scale 
in Fig.~\ref{fig:BO-potentials}b emphasizes the exponential long-range behavior of our result, $V(r)\sim \exp(-\sqrt{2} r/\xit)$ (see supplemental), which is sufficient
for experimentally relevant energy scales, while the Casimir term $\sim 1/r^3$ \cite{Schecter2014,Reichert2019PRB} affects only the already small tails of the potential.

In the limit $\eta\to\infty$ one finds the simple explicit form
\begin{equation}\label{eq:V-asymptot}
V(r)\Bigr\vert_{\eta \to \infty} = \frac{4}{3} \sqrt{2} g n_0^2 \xit + \frac{1}{2} g n_0^2 r \qquad \text{for } r \leq \pi \xit 
\end{equation}
where the potential is linear, corresponding to a constant attractive force acting between the impurities. This is because for strong repulsion the Bose gas is completely expelled in between the impurities as long as $r\lesssim \pi \bar \xi $ and the attractive force results only from the pressure of the Bose gas outside of the pair. This is further illustrated in the Supplemental Material .

In the BO limit of massive impurities, the effective interaction potential can be accurately obtained in QMC simulations. In Fig.~\ref{fig:BO-potentials} we compare our analytic predictions for repulsive impurity-boson coupling, $\eta > 0$, with QMC data and find excellent agreement within a few percent margin. Unless stated otherwise we used $N=100$ bosons in the QMC simulations. While lowest-order perturbative theory predicts the same interaction strength in repulsive ($\eta>0$) and attractive cases ($\eta <0$), non-perturbative approaches show that $V(r)$ is substantially stronger for attraction (see Fig.~\ref{fig:BO-potentials}b). This can be qualitatively understood as the maximal density defect produced is limited in the repulsive case by full depletion, while it is unlimited in the attractive case. This makes numerical calculations in the attractive case more challenging.

\paragraph{Bipolaron of finite impurity mass:--}

\begin{figure}[htb]
\centering
\includegraphics[width=0.47\textwidth]{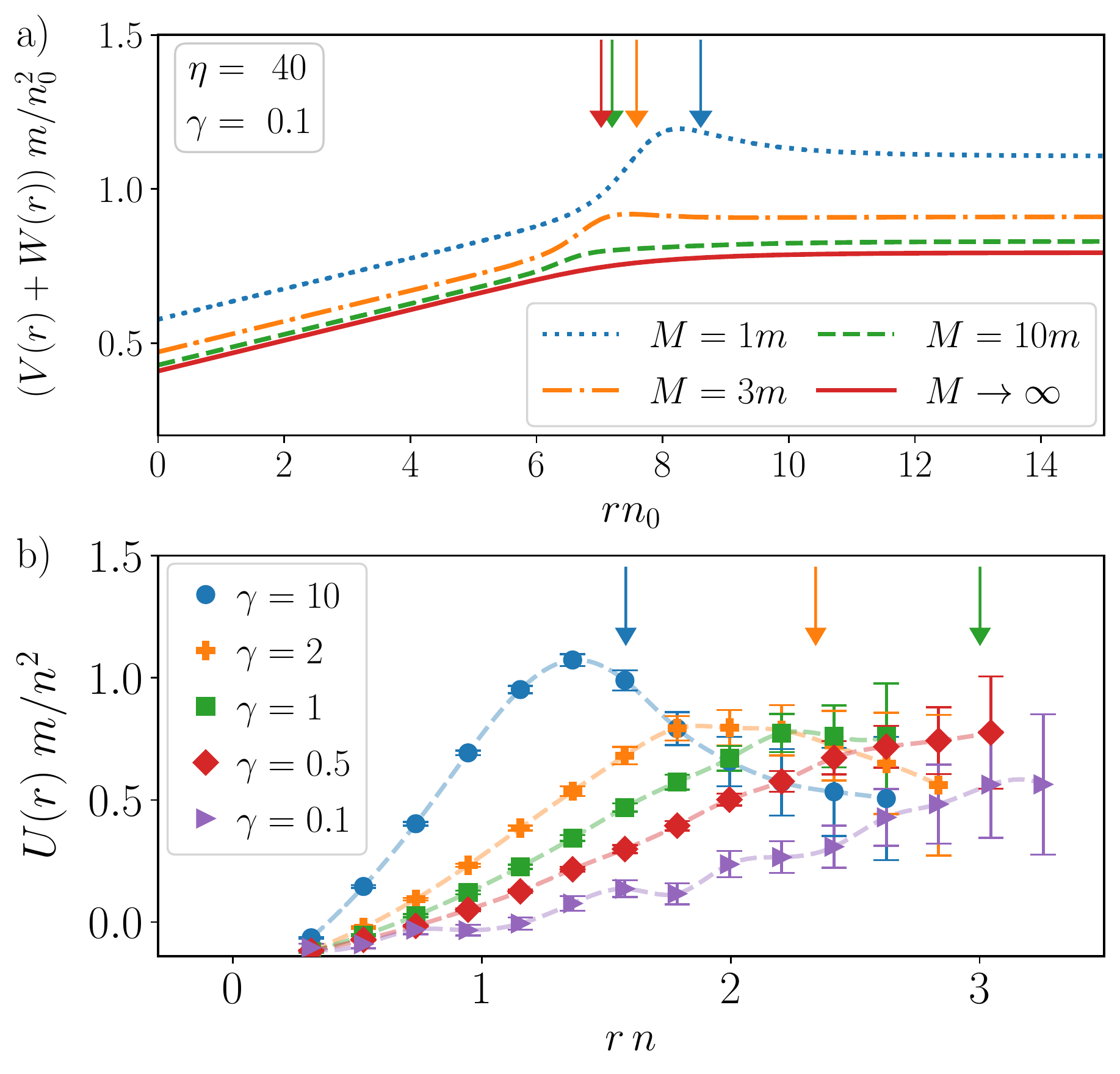}
\caption{Interaction potential for mobile impurities. (a) Mean-field potential including Born-Huang correction for different mass ratios. (b) Total interaction potential $U(r)$ from QMC simulations for the mass ratio $M=3m$, but $\eta \to \infty$ and different Lieb-Liniger parameters. Arrows point to analytical predictions of maxima $r_\text{max} = \pi \xit = \pi / \sqrt{2 \mred \mu}$, where we used the equation of state for $\mu$ from Bethe Ansatz \cite{Lang2017}. 
}
\label{fig:corrected-potential} 
\end{figure}

The BO approximation applies to infinitely heavy impurities {and} becomes increasingly inaccurate for light impurities.
The leading-order 
modification is the Born-Huang diagonal correction, $V(r)\to V(r)+W(r)$ \cite{Born1956,Fernandez1994}
\begin{align}
W(r) = \frac{1}{\mi}\int\, dx \, \left| \partial_r \, \phi_0(x) \right|^2, \label{eq:W_r}
\end{align}
where $\phi_0(x)$ is the mean-field wave function in the presence of two impurities at (fixed) distance $r$. $W(r)$ accounts for the dependence of the background-gas wave function on the impurity coordinates when calculating the impurity kinetic-energy. Including this term the approach is correct up to terms of order $\left(\mb/ \mi\right)^{3/2}$. Since the derivative of $\phi_0(x)$ with respect to $r$ is analytically complicated, we do not give an explicit expression for $W(r)$. In Fig.~\ref{fig:corrected-potential}a we plot the total potential for $\eta = 40$ and different characteristic mass ratios. Note that {the} finite impurity mass enters here in two ways, through the reduced healing length $\xit$ and by the Born-Huang term $W(r)$. A prominent feature is the emergence of a local 
maximum at distance $r_\textrm{max} \simeq \pi\xit$ when $W(r)$ is included. As discussed in the Supplemental Material this maximum appears only for strong impurity-boson coupling, i.e. if $\eta \gtrsim n_0 \xit$. Since for large values of $r$, $W(r)$ decays faster than $V(r)$, the total potential 
remains attractive at large distances.

\begin{figure}[htb]
\centering
\includegraphics[width=0.47\textwidth]{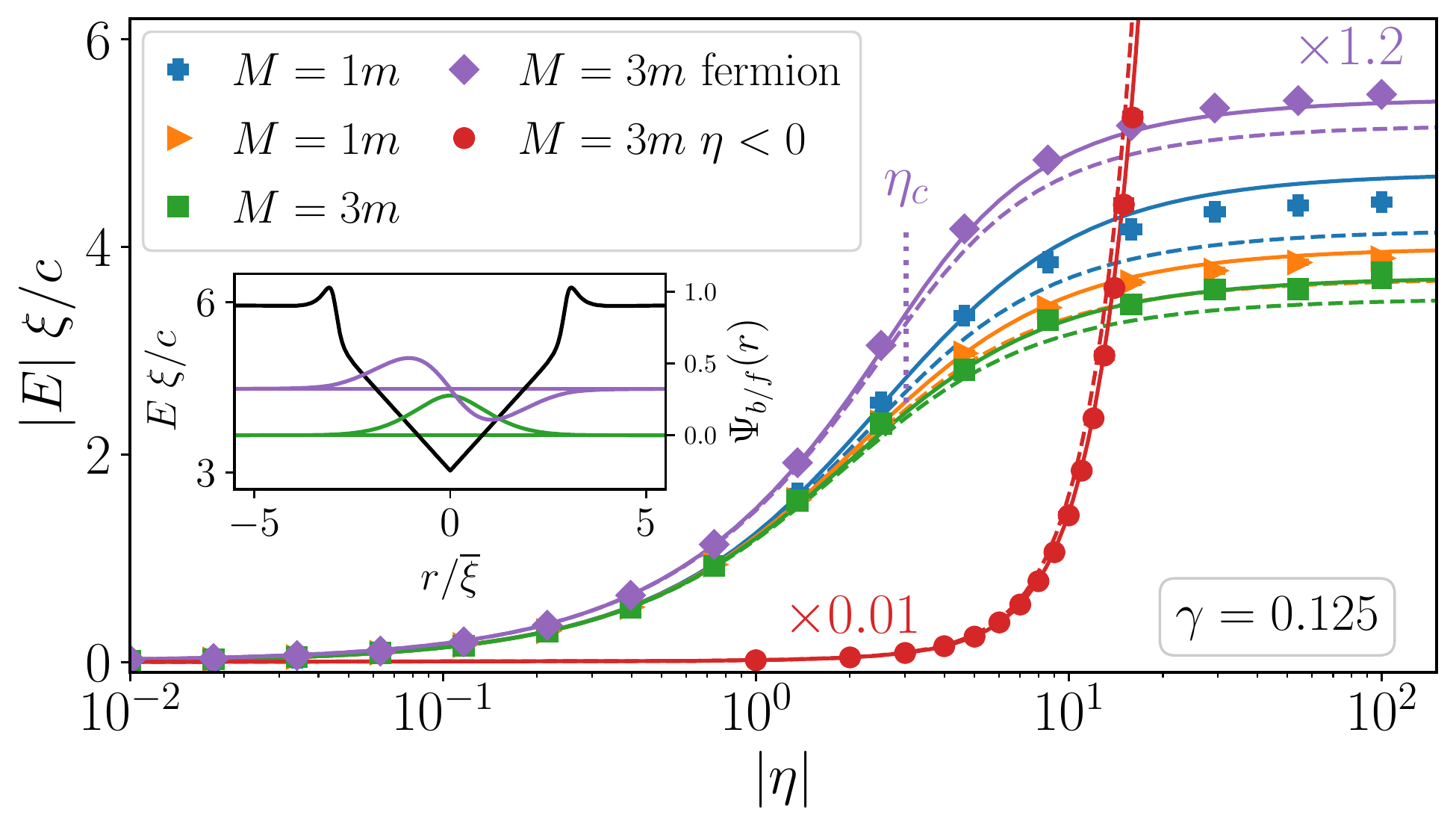}
\caption{Comparison of ground and first excited state energies of bipolarons with QMC results (dots) for different mass ratios $m/M$ and weak-to-moderate boson-boson coupling $\gamma =0.125$.
Dashed lines correspond to BO potential $V(r)$, solid lines include Born-Huang correction $V(r)+W(r)$. The red curves correspond to attractive impurity-boson interaction (scaled by $0.01$) and all others to repulsive. The purple line is the ground state energy for fermionic impurities (i.e. first excited bipolaron state, scaled by $1.2$), {where the vertical line marks the interaction strength $\eta_c \simeq 3$} {above which the two fermions form a bound state,} {calculated from the mean-field potential.} The inset ($\eta=30$ ; $M=3m$) illustrates the corresponding impurity wave functions in green (purple) for bosonic
(fermionic) impurities, as well as $V(r)+W(r)$ (black).}
\label{fig:energies}
\end{figure}

While for an infinite impurity mass, the interaction potential can be obtained directly in QMC simulations from the ground-state energy, its estimation is more delicate for finite values of $M$ and involves the impurity-impurity correlation function, $g_{ii}(x)$. Here 
the degrees of freedom of the gas are integrated out and $\sqrt{g_{ii}(x)}$ is interpreted as a wave function of the effective two-impurity Schr\"odinger equation. The effective potential is proportional to $(\sqrt{g_{ii}(x)})''/\sqrt{g_{ii}(x)}$, for details see Supplemental Materials. The large statistical noise arising from division by $\sqrt{g_{ii}(x)}$ does not allow to unambiguously identify a local potential maximum in a weakly interacting gas, $\gamma\ll 1$. The  maximum conjectured by the analytic theory is however clearly seen in the regime of strong interactions, $\gamma \gtrsim 1$, and although being outside the range of validity, its position is reasonably well predicted, see arrows in Fig.~\ref{fig:corrected-potential}b. Note that in the limit of a Tonks-Girardeau gas \cite{Girardeau1960}, $\gamma\to\infty$, the maxima coincide with the first maximum of Friedel oscillations \cite{Fuchs2007} at $n_0 r=1$ and in a super-Tonks-Girardeau gas would correspond to quasi-crystal lattice spacing. 
%
The attractive polaron interactions can lead to bound bipolaron states. In one dimension, {at least one} two-body bound states 
exist if the Fourier transform of the interaction potential at zero momentum is negative. We calculated the bipolaron energy of the lowest bound states for repulsive and attractive impurity-boson couplings with and without the Born-Huang corrections and compared the results to QMC simulations. While an attractive contact interaction only allows for a single bound state, here several ones are possible due to the finite extension of the effective potential. Note, however, that the first excited state of two bosonic impurities, mappable to the ground state of two fermions, becomes bound only above a critical interaction strength $\eta_c$. In Fig.~\ref{fig:energies} we plot the energies of the ground and first excited states of the bipolaron as a function of $\eta= \gib/g$ for a Bose gas with Lieb-Liniger parameter $\gamma=0.125$ for repulsive and attractive interactions.  Since the effective interaction potential is unbounded in the attractive case, much larger bipolaron energies are obtained for $\eta \to -\infty$. Once the Born-Huang corrections are included, an excellent quantitative agreement is found for mass ratios as small $\mi/m=3$. As shown in the Supplemental Material the predictions become less precise if the boson-boson interaction is increased, but even for $\gamma=1$, the discrepancy is below the few-percent level for $\eta \le 1$ and saturates below 15\% for $\eta \to \infty$. The bipolaron energies are in the same order as typical single polaron energies and in the strongly repulsive regime $\gib\gg g n_0 \overline{\xi}$ they are 
comparable to the energy of a dark soliton $E\sim \hbar n_0 c$. For the experimental data of Ref.~\cite{Catani2012}, where $n_0\approx 7 \mu $m$^{-1}$ and $c\approx 3.4$ mm/s, the latter corresponds to temperatures of $T=E/k_B\approx 240$ nK.

\paragraph{Conclusions:--}
We presented a detailed study of bipolarons and polaron-polaron interactions in ground-state 1D Bose gases. We have developed a semi analytical theory applicable for weakly interacting bosons and valid for arbitrarily strong impurity-boson interactions. As opposed to solid-state systems, where impurities couple only to collective excitations, the high compressibility of the 
Bose gas makes it necessary to take into account the action of the impurity to the quasi-condensate. This was done by expanding the quantum field of the bosons around a deformed quasi-condensate \cite{Jager2019}. In this way the density of phonons created by the impurities remains small also for strong impurity-boson couplings and phonon-phonon interactions can be disregarded. 
We derived the short-range potential from analytic mean-field solutions in BO approximation and found excellent agreement with QMC simulations. In the limit of strong impurity-boson interactions, $\gib/g \gg 1/\sqrt{2 \frac{m_\textrm{r}}{m}\gamma}$, the potential deviates substantially from the perturbative exponential form and attains a linear short-range dependence. When lowest-order corrections to the BO result are included, the potential becomes non-monotonic 
and attains a local maximum at a distance of $\pi\bar\xi$. As the interactions in the gas are made stronger, the height of the peak is increased and its position moves towards the first maximum of the Friedel oscillations. Comparison with QMC simulations shows that the analytic model provides a precise prediction for bipolaron energies for bosonic and fermionic impurities. Thus the mean-field description beyond the Froehlich model constitutes an excellent basis for the analysis of non-equilibrium and many-body properties of Bose polarons. Going away from equilibrium, e.g. by applying periodic drive or similar Floquet techniques will open new avenues to modify interactions of impurities mediated by a many-body environment with  applications to fields such as high-$T_c$ superconductivity and others. For this it is important to have tractable theoretical tools at hand. The application of our approach to the non-equilibrium physics of interacting polarons will be subject of future work.

\emph{Note:} After submission of this manuscript, we became aware of a recent related work on bipolarons in the limiting case of infinite impurity masses, using a different
approach~\cite{Volosniev2021}. The conclusions are in agreement with ours in this limit.

\emph{Acknowledgments}
We would like to thank Jonas Jager for fruitful discussions. M.W. and M.F. acknowledge financial support by the DFG through SFB/TR 185, Project No.277625399. M.W. acknowledges support from the Max Planck Graduate Center.
G.E.A. acknowledges financial support from the Spanish MINECO (FIS2017-84114-C2-1-P), and from the Secretaria d'Universitats i Recerca del Departament d'Empresa i Coneixement de la Generalitat de Catalunya within the ERDF Operational Program of Catalunya (project QuantumCat, Ref.~001-P-001644).
The authors thankfully acknowledge the computer resources at MareNostrum and the technical support provided by Barcelona Supercomputing Center (RES-FI-2020-3-0011).

\section{Supplemental Material}

\subsection{A. Quantum Monte Carlo simulations}

In order to obtain non-perturbative results we resort to the diffusion Monte Carlo method. It allows us to find the properties of a many-body system numerically, starting directly from the microscopic Hamiltonian which we take in the following form $(\hbar =1)$
\begin{align}  
\hat H
= &-\sum\limits_{i=1}^N\frac{1}{2m}\frac{\partial^2}{\partial x_i^2}
+\sum\limits_{i<j}^N g\delta(x_i-x_j) \notag \\
&-\sum\limits_{i=1}^{N_\textrm{I}}\frac{1}{2M}\frac{\partial^2}{\partial X_i^2} +\sum\limits_{i}^N\sum\limits_{j}^{N_\textrm{I}}\gib\delta(x_i-X_j),
\label{Eq:H}
\end{align}
where $x_i$ are positions of $N$ bosons and $X_j$ are positions of $N_\textrm{I}$ impurities. The simulation is performed in a box of size $L$ with periodic boundary conditions. 

The diffusion Monte Carlo (DMC) method is based on solving the Schrödinger equation in imaginary time. The contributions from excited states are exponentially suppressed for large propagation times and one is able to extract the ground-state energy exactly. The convergence is enhanced by using an importance-sampling technique in which the Schrödinger equation is solved for the product of the wave function and a guiding wave function which we take in the Jastrow pair product form
\begin{align}
&\psi_T(x_1,\cdots,x_N; X_1,\cdots,X_{N_\textrm{I})}\notag \\
&= \prod\limits_{i<j}^N\!f_\mathrm{BB}(x_i\!-\!x_j) 
\prod\limits_{i=1}^N\prod\limits_{j=1}^{N_\textrm{I}}\!f_\mathrm{BI}(x_i\!-\!X_j).
\end{align}
The Jastrow terms are chosen in such a way that when two bosons (or a boson and an impurity particle) meet, the delta-pseudopotential present in Hamiltonian~(\ref{Eq:H}) induces a kink in the wave function of a strength proportional to $g$ (or $\gib$). This is done by using at short distances the two-body scattering solution $f_{\mathrm{BB}}(x) = A_\mathrm{BB}\cos(k_\mathrm{BB}(x-B_\mathrm{BB}))$, $|x|<R_\mathrm{BB}$ which satisfies the Bethe-Peierls boundary condition, $f_{\mathrm{BB}}'(0)/f_{\mathrm{BB}}(0) = -1/a_\mathrm{BB}$ where $a_\mathrm{BB}$ is the $s$-wave scattering length. At larger distances this solution is matched with the long-range asymptotic obtained from hydrodynamic theory, $f_{\mathrm{BB}}(x) = \sin^{1/K_{\mathrm{BB}}}(\pi x/L)$ for $R_\mathrm{BB}<|x|<L/2$. The parameters $A_\mathrm{BB}$, $B_\mathrm{BB}$, $K_\mathrm{BB}$ are chosen from the Bethe-Peierls boundary condition at zero distance, the continuity conditions for the function and its first derivative at the matching distance $R_\mathrm{BB}$, while the periodic boundary condition $f_\mathrm{BB}'(L/2)=0$ is automatically satisfied. We consider a similar structure of the boson-impurity Jastrow terms $f_{\mathrm{BI}}(x)$.  Here the parameters $R_{BB}$, $R_{BI}$ and $K_{BI}$ are optimized in variational calculations by minimizing the variational energy.  $K_{\mathrm{BB}}(x)$ has the meaning of the Luttinger liquid parameter and its value is exactly known from Bethe ansatz.

The calculation of the effective impurity-impurity interaction potential is performed differently in the case of an impurity of an infinite or finite mass:

\paragraph{infinite impurity mass --}
In this case the interaction potential is calculated from the ground-state energies of the system with two impurities, $E_2$, at positions $X_1$ and $X_2$; and with no impurities $E_0$ according to
\begin{equation}
V(X_2-X_1) = E_2-E_0.
\end{equation}
The actual dependence on the relative distance between the impurities is obtained by repeating the calculations for different values of $X_2-X_1$. Each separate energy, $E_2$ and $E_0$, corresponds to the ground-state energy of the corresponding system and the DMC method allows exact calculation of such energies.

\paragraph{finite impurity mass --}
Here we calculate the impurity-impurity correlation function $g_{ii}(X_1-X_2)$ and interpret it as the square of wave function of the two-impurity solution. By doing so we effectively integrate all degrees of freedom associated with Bose particles. The  ground-state wave function $\psi_{ii}(X) = \sqrt{g_{ii}(X)}>0$ obtained in this way satisfies the two-impurity Schrödinger equation 
\begin{eqnarray}
-\frac{1}{2\mu^*}\psi''_{ii}(X)
+V(X)\psi_{ii}(X)
=E_{ii}\psi_{ii}(X),
\label{Eq:Schroedinger:ii}
\end{eqnarray}
where $\mu^*$ is the reduced effective polaron mass.
Note that we here assumed that the mediated impurity-impurity interaction can be described as an effective interaction $V(X)$ of two well-defined quasi-particles. This is justified for heavy impurities or weak coupling but ignores the possibility of an effective distance-dependent polaron mass. The excellent agreement between analytic calculations of the bipolaron binding energies based on this assumption and QMC data shows that this assumption is justified also for smaller mass ratios and in the strong interaction limit. Here $E_{ii} = E_2-E_0$ is the bipolaron binding energy, and $V(X)$ is the unknown effective interaction potential, which can be obtained from Eq.~(\ref{Eq:Schroedinger:ii}) according to
\begin{eqnarray}
V(X) = E_{ii} +\frac{1}{2\mu^*} \frac{(\sqrt{g_{ii}(X)})''}{\sqrt{g_{ii}(X)}}.
\label{Eq:Schroedinger_potential:ii}
\end{eqnarray}
That is, the bipolaron binding energy $E_{ii}$ provides a vertical offset in $V(X)$ while the reduced effective mass $\mu^*$ ``stretches'' the interaction potential vertically. The actual values of $E_{ii}$ and $\mu^*$ are not important for observing the predicted non-monotonous behavior, so for simplicity we use the bare impurity mass $\mu^*=M/2$.

\subsection{B. Single Bose polarons on a 1D ring}

The physics of single polarons in infinite one-dimensional Bose gases has been studied in Ref.~\cite{Jager2019} by assuming small quantum fluctuations on top of a deformed quasi condensate. We here illustrate the quantitative accuracy of this theory by comparison with quantum Monte Carlo simulations and extend the discussion by considering effects from the finite size of the 1D Bose gas.

Analytic solutions of the mean-field equations in the LLP frame were found and quantum fluctuations were taken into account in the lowest order by solving the corresponding Bogoliubov-de Gennes equations in Ref.~\cite{Jager2019}. While the agreement with QMC data for the polaron energy showed very good agreement, the predictions for the polaron mass $m^*$ had a different asymptotic for $\eta \to \infty$ as the QMC simulations. This discrepancy is a finite-size effect (QMC calculations have been performed with $N=50$ particles) which is much stronger in the polaron mass as compared to the energy. Indeed, if a finite number of bosons is considered on a ring then the polaron mass is bounded from above by the total mass of the bosons. 

\begin{figure}[htb]
\centering
\includegraphics[width=.9\linewidth]{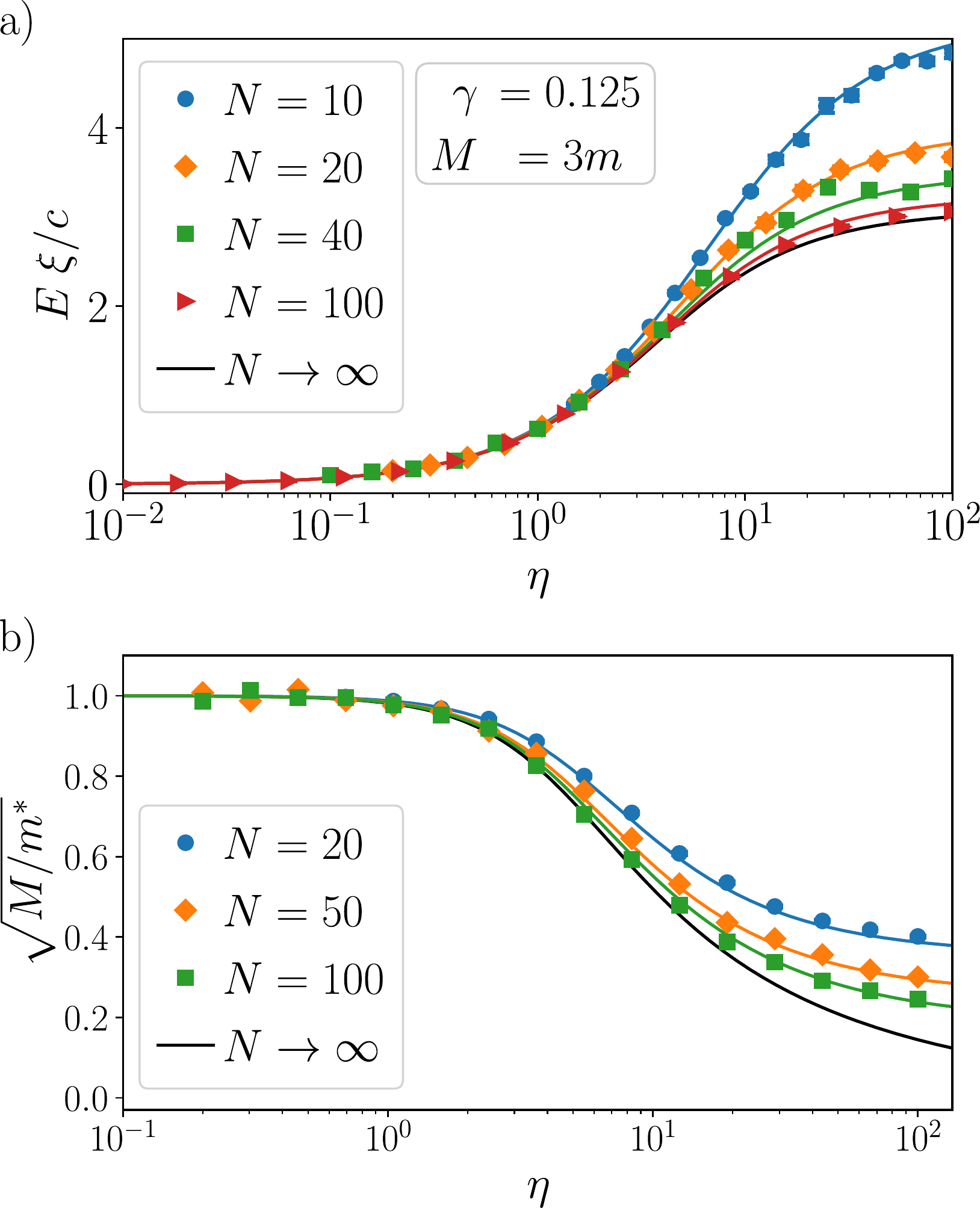}
\caption{Polaron (a) energy and (b) mass for a gas on a ring for different number of bosons. 
Symbols, QMC data;
lines, predictions of the analytic theory.
}
\label{fig:finite_size}
\end{figure}

Notably, the finite-size corrections can be correctly predicted in the mean-field theory by fixing the total number of particles rather than the condensate density far away from the impurity. In Fig.~\ref{fig:finite_size}(a) we have shown a comparison of the polaron energy for different system sizes and as a function of the impurity-boson coupling strength $\eta$ obtained from mean-field solutions and QMC simulations. One recognizes an excellent agreement. Increasing $\eta$, more and more bosons are expelled from the immediate surrounding of the impurity which leads to an enhancement of the boson density $n_0$ in the regions far away from the impurity. This in turn increases their mean-field energy $\propto gn_0^2 \xi $.

As can be seen from Fig.~\ref{fig:finite_size}(b) taking into account finite-size corrections also leads to a saturation of the effective polaron mass in the infinite coupling limit in excellent agreement with the predictions of QMC simulations.

We note that contrary to the 3D geometry where the finite-size correction due to periodic boundary conditions is an artifact, in 1D geometry periodic boundary conditions are physical and can be observed in cold gas experiments in a ring-shaped trap. Here qualitative new effects can arise if the relevant length scales, e.g. the reduced healing length $\xit$, become comparable to the system size $L$\cite{DeRosi2017}. 

\subsection{C. Details of the mean-field solution}

Using a Born-Oppenheimer and a mean-field approximation, one can derive the non-linear Schrödinger equation which determines the ground state of the Bose gas from the many body Hamiltonian (see Eq.~({\color{RubineRed}2}) in the main text):
\begin{align}
\Bigg[&-\frac{\partial_x^2}{2 \mred} \; + \; g \, |\phi_0(x)|^2 \; - \; \mu  \notag \\
 &+ \gib \Big( \delta(x-r/2) + \delta ( x+ r/2) \Big) \Bigg] \phi_0(x) = 0. \label{eq:NLSEQ}
\end{align}
This non-linear differential equation can be solved semi analytically using the Jacobi elliptic $\cd$ function \cite{Lawden1989}.
\begin{align}
\phi_0(x) = \sqrt{n_0}
\begin{cases}
 \, \sqrt{\frac{2\tilde{\nu}}{\nu+1}} \, \cd \left(\displaystyle{\frac{x}{ \xit \, \sqrt{1+\nu}}}\, , \, \nu \right)^{\pm 1} \quad
& |x| < r/2
\\ 
 \tanh \left(\displaystyle{\frac{|x|-x_0}{\sqrt{2} \,\xit}}\right)^{\pm 1} \quad &|x| > r/2,
\end{cases}
\label{eq:condensate_WF}
\end{align}
where the upper (lower) sign corresponds to $\eta = \gib/ g >0$ ($\eta <0$). The chemical potential in mean-field approximation is given by $\mu = g n_0$, where $n_0$
is the particle density far away from both impurities. In the thermodynamic limit one has $n_0= n= N /L $. The parameter $x_0$ is chosen such that the wave function is continuous at $x = \pm r/2$, while $\nu = \nu(r,\eta)$ is determined by the implicit relation in the main text and ensures the correct jump in the first derivative of the wave function, enforced by the double delta potential in Eq.~\eqref{eq:NLSEQ}.

\begin{figure}[htb]
    \centering
    \includegraphics[width=.9\linewidth]{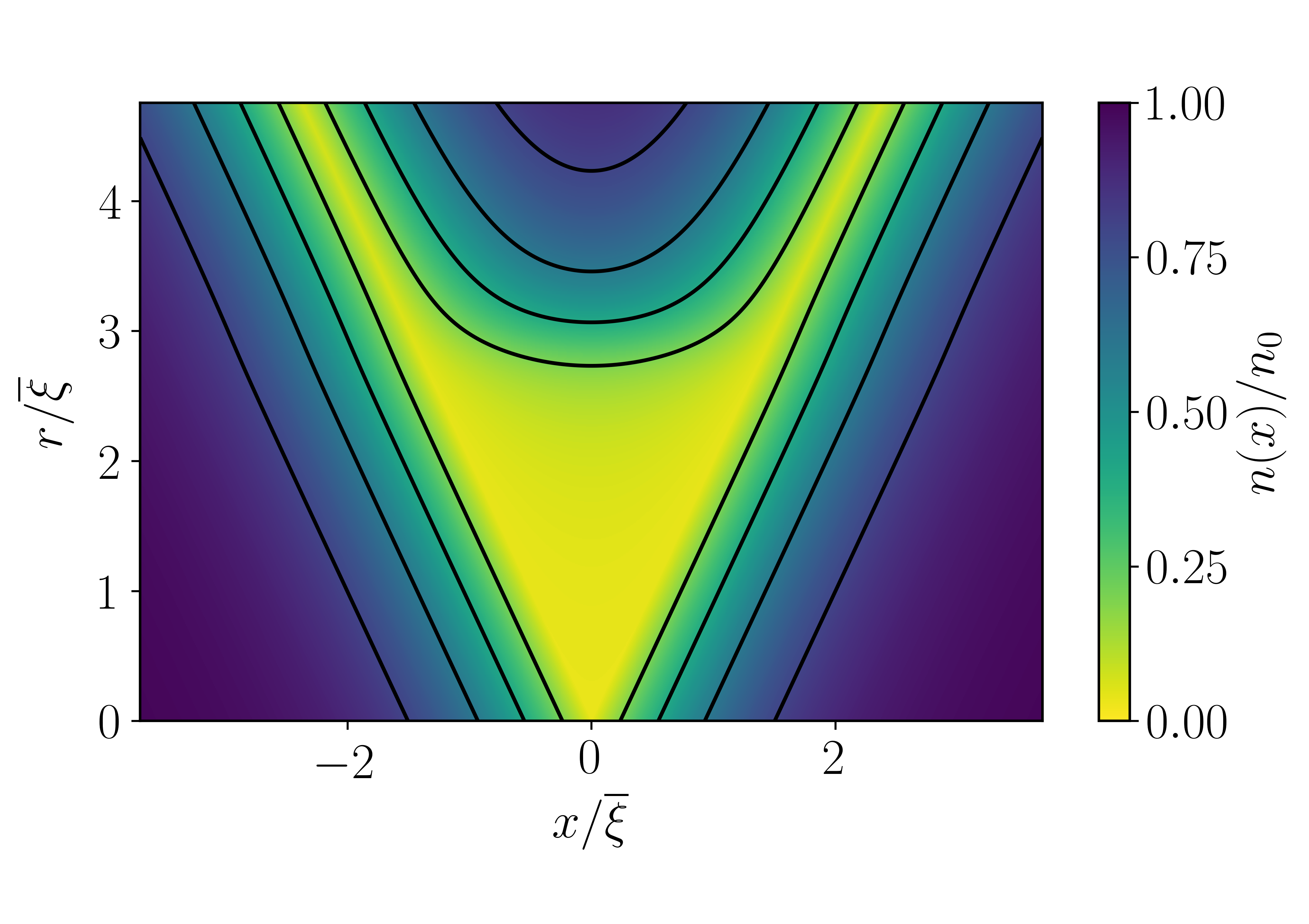}
    \caption{Bipolaron density $n(x)$ for different distances $r$ between the impurities and fixed interaction strength $\eta = 20 n_0 \xit$.}
    \label{fig:density}
\end{figure}

\begin{figure}[htb]
    \centering
    \includegraphics[width=.9  \linewidth]{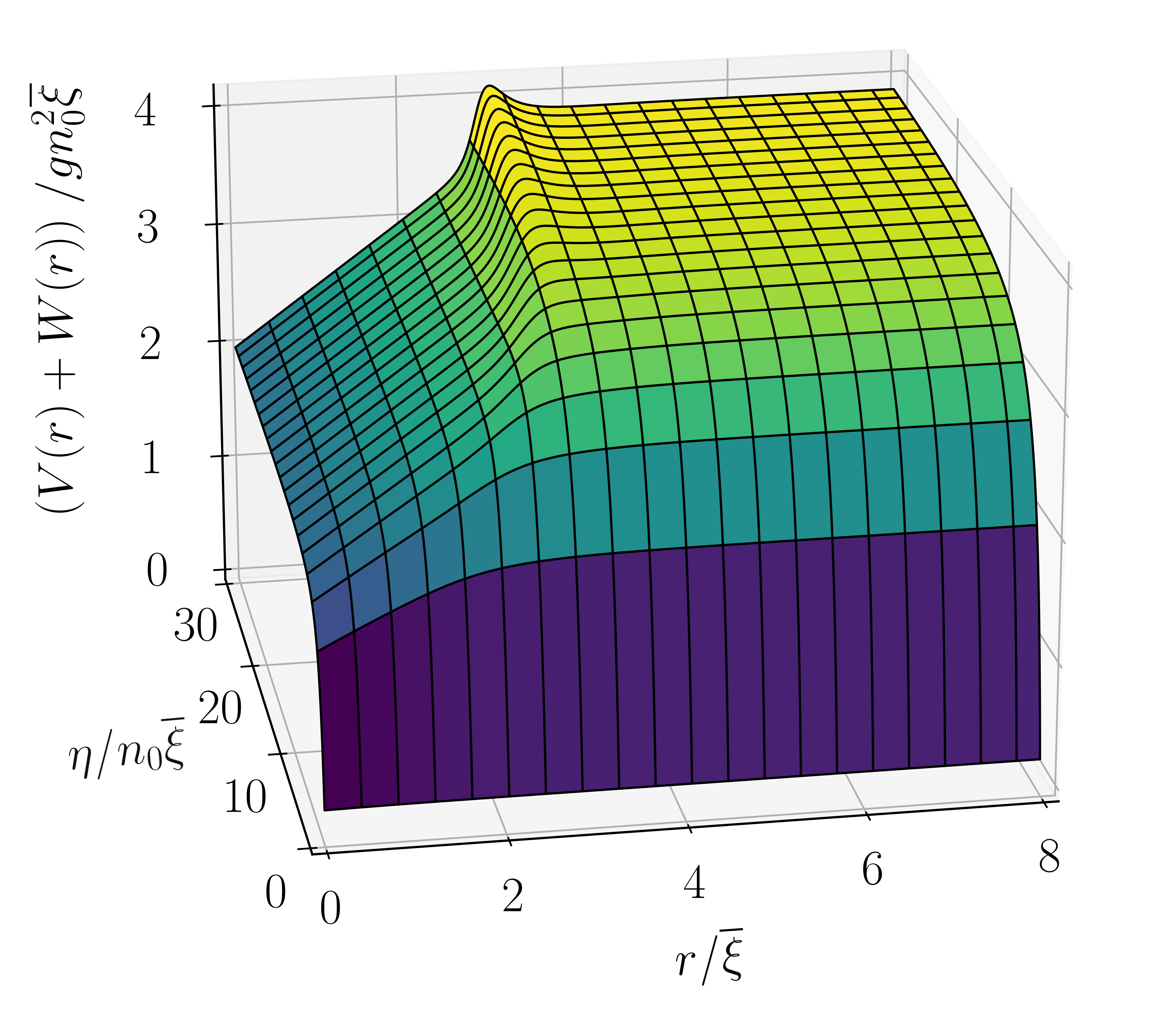}
    \caption{Effective potential including first-order correction to Born-Oppenheimer for mass ratio $M = 3 m$}
    \label{fig:3D_Potential}
\end{figure}

The density is plotted for two strongly interacting impurities, $\eta = 20 n_0 \xit$, in Fig.~\ref{fig:density}. For two impurities that are far away from each other, {$r \gg \xit = 1/\sqrt{2 g n_0 m_r}$}, the density profile is well approximated by the wave functions of two separated impurities \cite{Jager2019}. If the distance decreases to about $r\simeq \pi \xit $ we find a sudden reduction of the boson density in between the impurities and for a smaller distance 
all bosons are pushed out of this space. The distance $r/\xit \simeq \pi$ at which this depletion takes place is only an approximation for finitely repulsive impurities, but is exact for infinite repulsion $\eta \to \infty$. The parameter $\nu$ is in this limit determined by
\begin{alignat}{2}
\nu\, \Bigr\vert_{\eta \to \infty} &= 0 && \text{for} \; r \leq \pi \xit \notag \\
K(\nu) \, \sqrt{1+\nu} \, \Bigr\vert_{\eta \to \infty}&= \frac{r}{2 \xit} \quad && \text{for} \; r > \pi \xit ,
\end{alignat}
where $K(\nu)$ is the complete elliptic integral of the first kind. Inserting this into the mean-field wave function, Eq.~\eqref{eq:condensate_WF}, shows that the condensate density is totally depleted in between the two impurities for $r \leq \pi \xit$. In the next step the Born-Oppenheimer potential can be calculated as the energy of the mean-field wave function Eq.~\eqref{eq:condensate_WF}.
\begin{align}
V(r) = g n_0^2 L - \frac{g}{2} \int_{-L/2}^{L/2} \; \phi_0(x)^4 \, dx - E_0(\gib=0)
\end{align} 
where the extensive mean-field ground state energy $E_0(\gib=0)$ of the Bose gas without impurities is subtracted. In general this evaluates to Eq.~({\color{RubineRed}3}) in the main text, but the expression can again be simplified for infinitely repulsive impurities, especially the short-range potential in the limit $\eta\to\infty$ reads
\begin{align} \label{eq:V_large_eta}
&V(r) \Bigr\vert_{\eta \to \infty}   \notag \\
=&g n_0^2
\begin{cases}
\frac{4}{3} \sqrt{2} \xit + \frac{1}{2} r \quad \quad \text{for } r \leq \pi \xit \\
\frac{4}{3} \xit \left[\sqrt{2} + \displaystyle{\frac{E(\nu)}{\sqrt{1+\nu}}} \right]
+ r \; \displaystyle{\frac{-5+2 \nu +3 \nu^2}{6 (1+\nu)^2}} \quad \text{else,}
\end{cases} 
\end{align}
where $E(\nu)$ is the complete elliptic integral of the second kind. The linear short-range potential is caused by the bosons being expelled from the space in between the impurities. The bosons outside of this space therefore push the impurities towards each other with a constant force, which is equivalent to the linear potential. The potential starts to deviate from the linear slope as soon as the particle density in between the impurities increases. 
For a large distance $r \gg \xit$ between the two impurities we find a exponentially decaying potential which is for repulsive interaction given by
\begin{align}
&V(r)\Bigr\vert_{\eta>0\, , \, r\gg \xit}= 2E_1  \notag \\
&- 128 \sqrt{2}\; g n_0^2 \xit\; \frac{ n_0^2 \xit^2}{\eta^2} \left( \sqrt{1+\eta^2 / 8 n_0^2 \xit^2} -1 \right)^2 \; e^{-\sqrt{2} r /\xit},
\end{align}
where $E_1$ is the constant mean-field energy of a single polaron \cite{Jager2019}.

The first order Born-Oppenheimer correction
\begin{align}
 W(r) = \frac{1}{\mi} \int \; dx \left| \partial_r \phi_0(x) \right|^2\;,
\end{align}
accounts for the contribution of the condensate wave function, which depends on the impurity coordinate, when calculating the kinetic energy of the relative motion of the impurities \cite{Born1956,Fernandez1994}. Since the derivative of the wave function $\phi_0(x)$ \eqref{eq:condensate_WF} with respect to the distance $r$ is analytically difficult, we do not give an explicit expression.
A very prominent effect of this correction is the emergence of a local maximum in the two-particle potential Fig.~\ref{fig:3D_Potential}. It appears only for strongly interacting impurities $\eta \gg n_0 \xit$, when the deformation of the quasi-condensate is substantial. For small values of $r$, where the quasi-condensate in between the impurities is depleted, the correction is only a small constant. For a distance $r\simeq \pi \xit$ the density in between becomes nonzero again, and grows rapidly when further increasing the distance $r$, leading to a large correction. When the impurities are far apart, the contribution of the Bose gas to the kinetic energy of impurities is small again, which results in a local maximum of $W(r)$ at the intermediate distance. Since the relative kinetic energy of the impurities scales inversely with the impurity mass $M$, the potential correction also decreases with increasing value of $M$, which is shown in Fig.3a of the main part.

\subsection{D. Bipolaron binding energy for attractive interactions}

\begin{figure}[htb]
\centering
\includegraphics[width=0.45\textwidth]{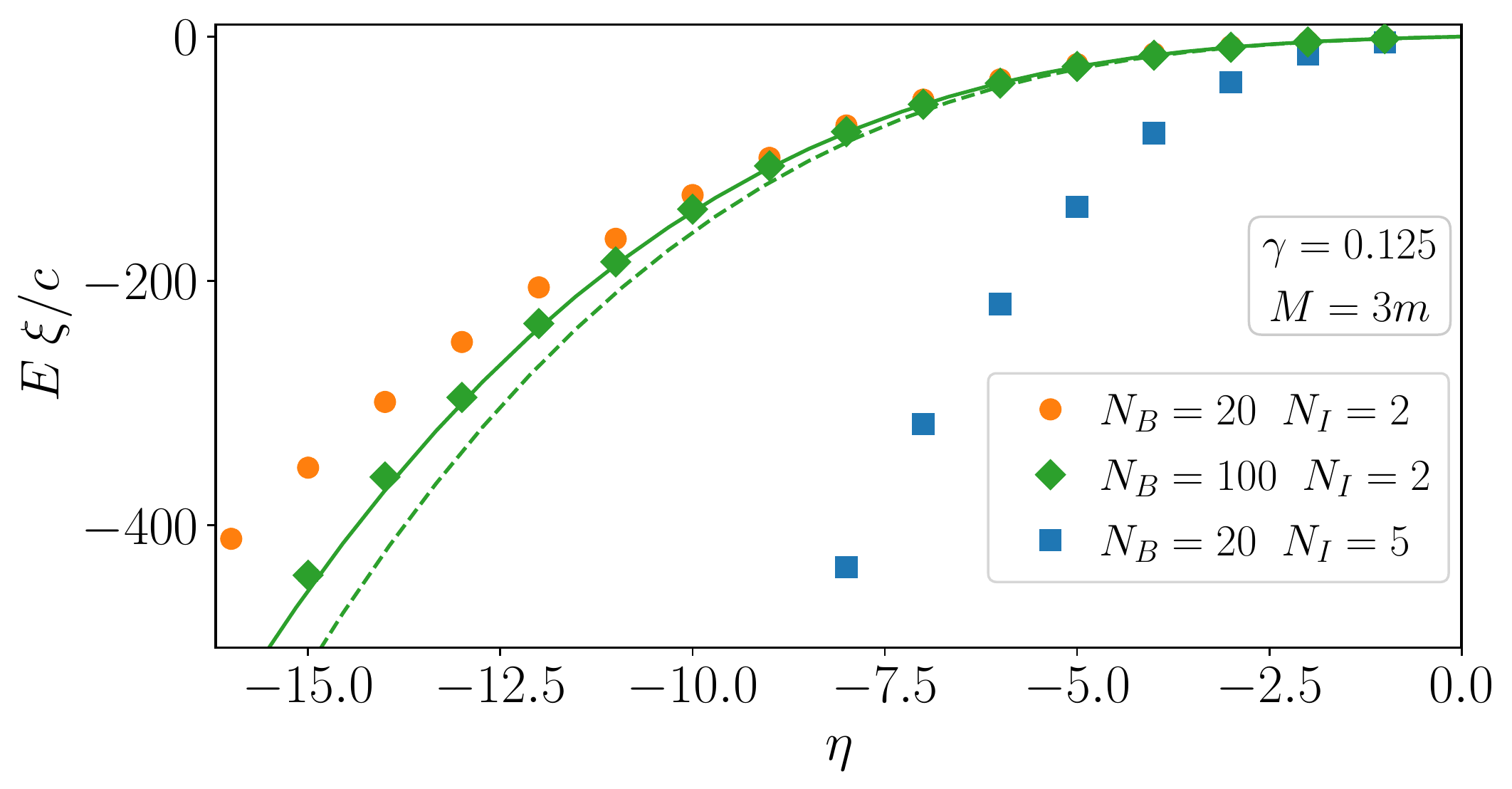}
\caption{QMC ground state energy of a system containing $N_\text{B}$ bosons and $N_\text{I}$ impurities interacting attractively with each other. The solid (dashed) line is the MF prediction of the bipolaron energy including (excluding) the first order BO correction in the thermodynamic limit. }
\label{fig:energies_attr}
\end{figure}

For repulsive impurity-boson interactions the two-particle potential approaches an asymptotic value with the linear short-distance behavior given in Eq.~\eqref{eq:V_large_eta}, when $\eta \to\infty$. The bipolaron binding energy thus saturates as can be seen in Fig.~{\color{RubineRed}4} of the main text. This is different in the attractive case, $\eta <0$, since more and more bosons can be pulled towards the impurities. Thus the bipolaron binding energy grows unlimited for $\vert \eta\vert \to \infty$. This can be seen in Fig.~\ref{fig:energies_attr}, where we have plotted the theoretical predictions resulting from the effective potential with and without BO corrections and compared them to QMC simulations. As one can see, the results are in a perfect quantitative agreement. Note that the discrepancy between $N_B=20$ and $N_B= 100$ bosons is a finite size effect {which becomes sizable for strong interactions, $\vert \eta\vert \ge 10$}.  It has a direct physical relevance in a one dimensional system, since periodic boundary conditions can be realized in experiments using a system on a ring trap. The QMC data for $N_I = 5$ impurities illustrate that a multi-impurity bound state can form in the case of an attractive impurity-boson interaction with a rapidly increasing binding energy.

\subsection{E. Effect of strong boson-boson interactions}

\begin{figure}[htb]
\centering
\includegraphics[width=0.45\textwidth]{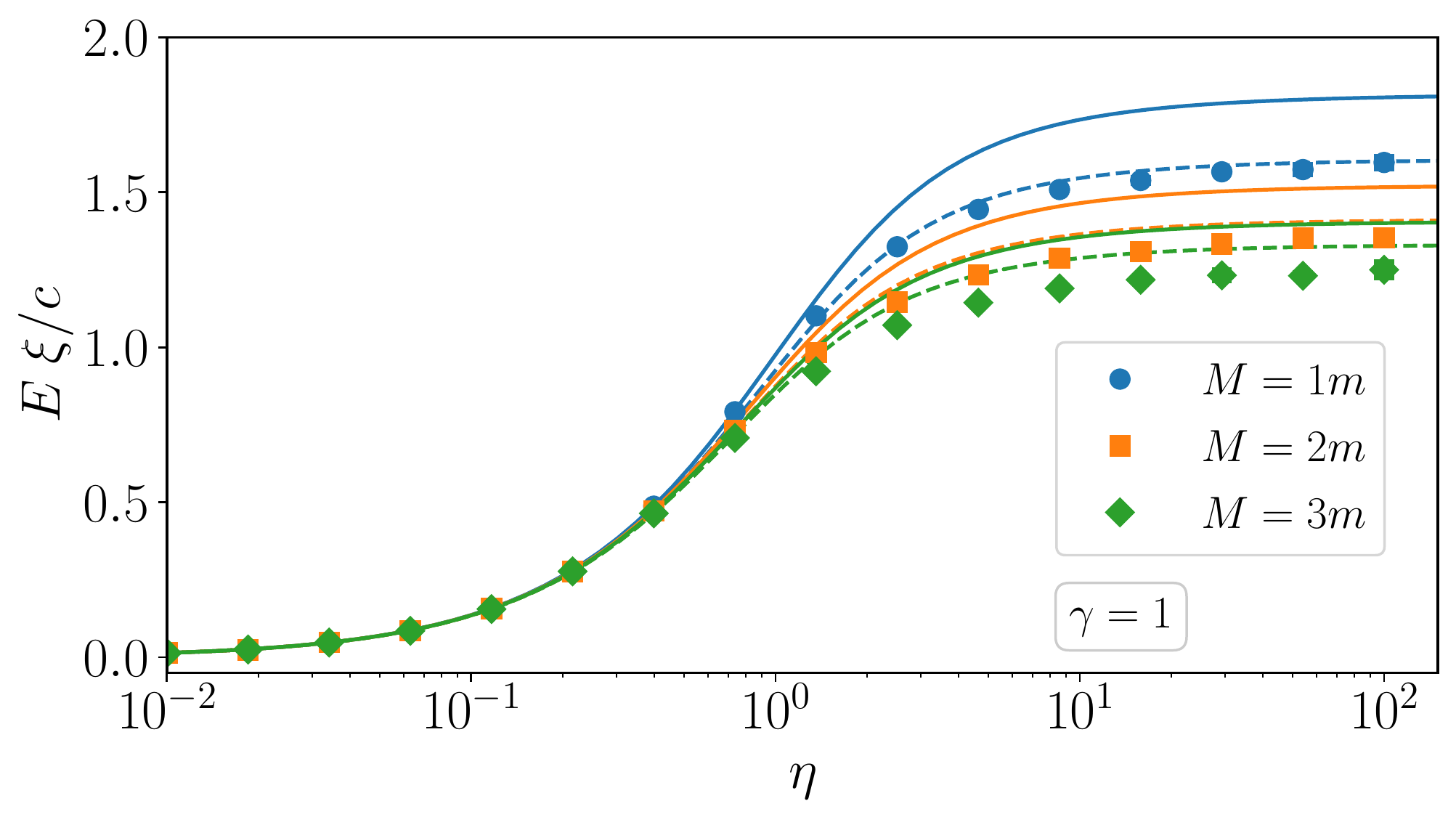}
\caption{Comparison of ground state energy of bound bipolaron and QMC for different mass ratios $m/M$ and strong boson-boson coupling $\gamma=1$. Solid (dashed) lines correspond to theoretical predictions with (without) BO corrections and the dots denote numerical values obtained from QMC simulations.}
\label{fig:energies_large_gamma}
\end{figure}

The semi-analytic theory presented in the main text is valid for arbitrarily large impurity couplings but is limited to weak boson-boson interactions, corresponding to small Lieb Liniger parameter $\gamma$. In Fig.~\ref{fig:energies_large_gamma} we have plotted the bipolaron binding energies for strong boson-boson interactions ($\gamma=1$) for different impurity-boson mass ratios as function of the impurity coupling. While there are increasing deviations from exact numerical QMC values, the agreement is still rather good for $\eta \lesssim 1$. 

\bibliographystyle{apsrev4-2} 

\bibliography{library}

\end{document}